%% file: paper.tex
\documentclass[10pt,conference]{IEEEtran}
\IEEEoverridecommandlockouts

\usepackage{stmaryrd}
\usepackage{mathpartir}
\usepackage{amssymb}
\usepackage{listings}
\usepackage{tikz}
\usepackage{hyperref}
\usepackage{setspace}
\usepackage{etoolbox}
\usepackage[capitalise,nameinlink]{cleveref}
\usepackage[linesnumbered,ruled]{algorithm2e}

\usetikzlibrary{shapes.geometric, arrows.meta, positioning, fit, calc}

\crefformat{section}{#2\S{}#1#3}
\crefname{algocf}{Alg.}{Algorithms}

\definecolor{RoyalBlue}{HTML}{0071BC}
\definecolor{ForestGreen}{HTML}{009B55}
\definecolor{ACMPurple}{cmyk}{0.55,1,0,0.15}
\definecolor{ACMDarkBlue}{cmyk}{1,0.58,0,0.21}
\definecolor{backcolor}{rgb}{0.95,0.95,0.95}

\hypersetup{%
  linktocpage=true, pdfstartview=FitV,
  breaklinks=true, pageanchor=true, pdfpagemode=UseOutlines,
  plainpages=false, bookmarksnumbered, bookmarksopen=true, bookmarksopenlevel=3,
  hypertexnames=true, pdfhighlight=/O,
  colorlinks=true,
  linkcolor=ACMPurple,
  citecolor=ACMPurple,
  urlcolor=ACMDarkBlue
}

\lstdefinestyle{mystyle}{
  backgroundcolor=\color{backcolor},
  keywordstyle=\color{RoyalBlue},
  commentstyle=\color{ForestGreen},
  basicstyle=\linespread{0.8}\ttfamily\footnotesize,
  breakatwhitespace=false,
  breaklines=true,
  captionpos=b,
  keepspaces=true,
  numbers=none,
  numbersep=1pt,
  numberstyle=\scriptsize,
  showspaces=false,
  showstringspaces=false,
  escapechar=$,
}
\lstdefinelanguage{Rust}{
  keywords={as,break,const,continue,crate,else,enum,extern,false,fn,for,if,
  impl,in,let,loop,match,mod,move,mut,pub,ref,return,self,Self,static,struct,
  super,trait,true,type,unsafe,use,where,while,async,await,dyn,abstract,become,
  box,do,final,macro,override,priv,typeof,unsized,virtual,yield,try,union,raw},
  morecomment=[l]{//},
  morecomment=[s]{/*}{*/},
}

\newcommand{\embox}[1]{{\mbox{\textit{#1}}}}
\newcommand{\C}[1]{{\small\ensuremath{\tt#1}}}
\newcommand{\CT}[1]{{\small\texttt{#1}}}
\newcommand{\ostdin}{{\sf stdin}}
\newcommand{\ostdout}{{\sf stdout}}
\newcommand{\ostderr}{{\sf stderr}}
\newcommand{\ofile}{{\sf file}}
\newcommand{\opipe}{{\sf pipe}}
\newcommand{\cread}{{\sf read}}
\newcommand{\cbufread}{{\sf bufread}}
\newcommand{\cwrite}{{\sf write}}
\newcommand{\cseek}{{\sf seek}}
\newcommand{\cclose}{{\sf close}}

\newcommand{\loc}[1]{{\it Loc}(#1)}
\newcommand{\struct}[1]{{\it Struct}(#1)}
\newcommand{\oriset}[1]{\llbracket#1\rrbracket_o}
\newcommand{\capset}[1]{\llbracket#1\rrbracket_c}
\newcommand{\boundof}[1]{{\it bound}(#1)}

\newcommand{\tread}{{\tt Read}}
\newcommand{\tbufread}{{\tt BufRead}}
\newcommand{\twrite}{{\tt Write}}
\newcommand{\tseek}{{\tt Seek}}
\newcommand{\tstdin}{{\tt Stdin}}
\newcommand{\tstdout}{{\tt Stdout}}
\newcommand{\tstderr}{{\tt Stderr}}
\newcommand{\tfile}{{\tt File}}
\newcommand{\tbufreader}{{\tt BufReader\texttt{<}File\texttt{>}}}
\newcommand{\tbufwriter}{{\tt BufWriter\texttt{<}File\texttt{>}}}
\newcommand{\tchild}{{\tt Child}}
\newcommand{\tptr}[1]{{\tt *mut}\ #1}
\newcommand{\tbox}[1]{{\tt Box\langle dyn}\ #1\rangle}
\newcommand{\tdynptr}[1]{{\tt *mut\ dyn}\ #1}
\newcommand{\timpl}[1]{{\tt impl}\ #1}
\newcommand{\eptr}[1]{{\tt\&raw\ mut}\ #1}

\begin{document}

\title{Forcrat: Automatic I/O API Translation from C to Rust via Origin and Capability Analysis}

\author{
  \IEEEauthorblockN{
    Jaemin Hong\IEEEauthorrefmark{1}\IEEEauthorrefmark{2} and Sukyoung Ryu\IEEEauthorrefmark{3}}
  \IEEEauthorblockA{\IEEEauthorrefmark{1}
  \textit{Institute of Information and Electronics}, \textit{KAIST}, Daejeon, South Korea}
  \IEEEauthorblockA{\IEEEauthorrefmark{2}
  \textit{Department of Computer Science}, \textit{Yale University}, New Haven,
  Connecticut, USA}
  \IEEEauthorblockA{\IEEEauthorrefmark{3}
  \textit{School of Computing}, \textit{KAIST}, Daejeon, South Korea}
  \IEEEauthorblockA{jaemin.hong@kaist.ac.kr, sryu.cs@kaist.ac.kr}
}

\maketitle

\begin{abstract}

  Translating C to Rust is a promising way to enhance the reliability of legacy
  system programs.
  Although the industry has developed an automatic C-to-Rust translator, C2Rust,
  its translation remains unsatisfactory.
  One major reason is that C2Rust retains C standard library (libc) function
  calls instead of replacing them with functions from the Rust standard library
  (Rust std).
  However, little work has been done on replacing library functions in
  C2Rust-generated code.
  In this work, we focus on replacing the I/O API, an important subset of
  library functions.
  This poses challenges due to the semantically different designs of I/O APIs in
  libc and Rust std.
  First, the two APIs offer different sets of types that represent the
  \emph{origins} (e.g., standard input, files) and \emph{capabilities} (e.g.,
  read, write) of streams used for I/O.
  Second, they use different error-checking mechanisms: libc uses internal
  indicators, while Rust std uses return values.
  To address these challenges, we propose two static analysis techniques,
  \emph{origin and capability analysis} and \emph{error source analysis}, and
  use their results to replace the I/O API.
  Our evaluation shows that the proposed approach is
  (1) correct, with all 32 programs that have test suites passing the tests
  after transformation,
  (2) efficient, analyzing and transforming 422k LOC in 14 seconds, and
  (3) widely applicable, replacing 82\% of I/O API calls.

\end{abstract}


\input{intro}
\input{background}
\input{analysis}
\input{transformation}
\input{evaluation}
\input{related}
\input{conclusion}

\section*{Acknowledgment}

This material is based upon work supported in part by
the Defense Advanced Research Projects Agency (DARPA) under Agreement No.
HR00112590130,
the National Research Foundation of Korea (NRF) (2022R1A2C2003660 and
2021R1A5A1021944),
Institute of Information \& Communications Technology Planning \& Evaluation
(IITP) grant funded by the Korea government (MSIT) (2024-00337703),
the KAIST Jang Young Sil Fellow Program,
and Samsung Electronics Co., Ltd. (G01240469).
Any opinions, findings, and conclusions or recommendations expressed in this
material are those of the authors and do not necessarily reflect the views of
the funding agencies.

\bibliographystyle{IEEEtran}
\bibliography{references}

\end{document}

%% file: intro.tex
\section{Introduction}
\label{sec:intro}

Translating C to Rust is a promising approach to enhancing the reliability of
legacy system programs.
Since C lacks language-level mechanisms to ensure memory safety, legacy software
written in C has suffered from critical security vulnerabilities caused by
memory bugs~\cite{chen2011linux, msrcblog}.
Rust is a modern systems programming language that addresses this issue by
providing a type system that guarantees memory safety at compile
time~\cite{matsakis2014rust, jung2017rustbelt}.
Thus, translating legacy code to Rust enables developers to detect previously
unknown bugs and reduce the risk of introducing new bugs~\cite{rust-curl,
li2024rust}.

Since code translation is laborious and error-prone when done manually, the
industry has developed an automatic C-to-Rust translator named
C2Rust~\cite{c2rust}.
It can process various real-world C codebases, producing Rust code that is
syntactically valid and semantically equivalent to the original program.
Software companies and open-source projects have used C2Rust to translate their
code~\cite{c2rust-huawei, zebra, qcms}.

Unfortunately, the translation produced by C2Rust is unsatisfactory because the
memory safety of the generated Rust code cannot be ensured by the Rust compiler,
which contradicts the goal of the translation.
The main reason is the use of C standard library (libc) functions through Rust's
foreign function interface.
C2Rust retains each libc function call during translation, instead of replacing
it with an equivalent function from the Rust standard library (Rust std).
However, foreign functions are not checked by the Rust compiler, potentially
compromising the memory safety of the entire program.

Therefore, replacing libc functions in C2Rust-generated code with proper Rust
std functions is an important problem for ensuring the safety of system programs
through translation.
Nevertheless, little work has been done on library replacements.
Most studies aimed at improving C2Rust-generated code have focused on language
features other than library functions, such as
pointers~\cite{emre2021translating, emre2023aliasing, zhang2023ownership,
wu2025genc2rust}, unions~\cite{hong2024tag}, and output
parameters~\cite{hong2024dont}.
One exception is the work by Hong and Ryu on the Lock
API~\cite{hong2023concrat}, but it addresses only a small subset of library
functions used in concurrent programs.

In this work, we aim to replace the I/O API in C2Rust-generated code.
The I/O API is one of the most important subsets of library functions, as almost
all C programs use it to interact with users (through terminals), files, or even
subprocesses, especially given C's primary role as a systems programming
language.
Specifically, we focus on replacing the I/O API provided by the \C{stdio.h}
header file in C with the one provided by Rust std.

The task of replacing the I/O API poses challenges due to the different designs
of the two APIs.
The libraries of the two languages differ not only syntactically, such as in
function names and argument order, where the conversion mapping can be
constructed manually or using existing mapping mining
techniques~\cite{zhong2010mining, nguyen2014statistical, meng2012history}.
They also differ semantically, primarily in two aspects: (1) types and (2) error
checking.

First, the two APIs provide different sets of types to perform I/O operations.
In libc, the type of a \emph{stream}---a target of each I/O operation that the
program reads from or writes to---is always \C{FILE*}.
In contrast, Rust std distinguishes the \emph{origins}, from which streams
originate, by defining multiple types such as \C{Stdin} and \C{File},
and the \emph{capabilities}, which regulate the operations that can be performed
on streams, through types such as \C{Read} and \C{Write}.
Therefore, replacing the I/O API requires determining the origin and capability
of each stream to assign the correct type to it.

Second, the two APIs offer different error checking mechanisms.
Since I/O operations may fail for various reasons,
%
%
the APIs provide ways to check whether each operation succeeded.
In libc, each operation silently sets an error indicator on the stream, which
the program can later inspect by calling functions like \C{ferror}.
This allows programs to check for errors at a different location (e.g., in
another function) from where the I/O operation was performed.
In contrast, each function in Rust std directly reports whether an error
occurred by returning a \C{Result} value, which is either \C{Ok} or \C{Err}.
As a result, replacing the I/O API requires propagating the directly returned error
indication values to the locations where the original program checks them.

\begin{figure}[t]
\begin{tikzpicture}[
  node distance=0.2cm and 0.4cm,
  auto,
  data/.style={
    rectangle,
    align=center,
    minimum height=1.8em
  },
  comp/.style={
    rectangle,
    draw,
    align=center,
    minimum height=1.8em
  },
  line/.style={draw, -Latex}
]
  \node [data, minimum height=1em] (ccode) {\scriptsize C Code};
  \node [comp, below=1.1em of ccode] (c2rust) {\scriptsize C2Rust};
  \node [data, right=of c2rust] (rustcode1) {\scriptsize Rust code\\[-0.6em]\scriptsize (libc I/O)};
  \node [comp, below=of rustcode1] (analyzer)
  {
    \scriptsize Origin and capability analysis (\cref{sec:analysis:oricap})\\[-0.6em]
    \scriptsize Error source analysis (\cref{sec:analysis:error})
  };
  \node [data, right=of analyzer] (result)
  {
    \scriptsize Streams' origins \& capabilities\\[-0.6em]
    \scriptsize Indicator-checking calls' sources
  };
  \node [comp, above=of result] (transformer) {\scriptsize Program transformation (\cref{sec:transformation})};
  \node [data, minimum height=1em, above=1.3em of transformer] (rustcode2)
    {\scriptsize Rust code\\[-0.6em]\scriptsize (Rust std I/O)};
  \path [line] (ccode) -- (c2rust);
  \path [line] (c2rust) -- (rustcode1);
  \path [line] (rustcode1) -- (transformer);
  \path [line] (transformer) -- ($(rustcode2.south)!0.1!(rustcode2.north)$);
  \path [line] ($(rustcode1.south)!0.1!(rustcode1.north)$) -- (analyzer);
  \path [line] (analyzer) -- (result);
  \path [line] ($(result.south)!0.8!(result.north)$) -- (transformer);
  \node[draw, inner xsep=3pt, inner ysep=6pt, fit=(c2rust) (rustcode1) (transformer) (analyzer) (result)] (largebox) {};
\end{tikzpicture}
  \vspace{-1em}
\caption{The workflow of the proposed approach}
  \vspace{-1.5em}
\label{fig:workflow}
\end{figure}
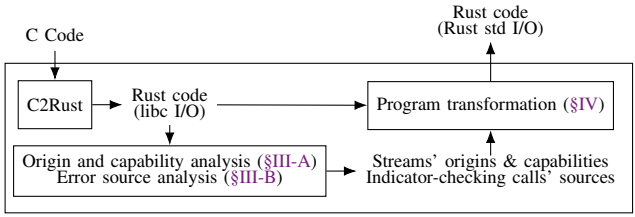

To address these challenges, we propose a static-analysis-based approach,
depicted in \cref{fig:workflow}.
Once C2Rust generates Rust code that still uses the libc I/O API, we perform two
static analyses on this code:
(1) \emph{origin and capability analysis}, a flow-insensitive analysis that
collects constraints on the origins and capabilities of streams and solves them,
and (2) \emph{error source analysis}, an interprocedural dataflow analysis that
identifies the API function calls responsible for setting indicators read by
each indicator-checking call.
We then transform the program to replace the I/O API, using the analysis results
to determine the type of each stream and to propagate errors.

Overall, our contributions are as follows:
\begin{itemize}
  \item
    We propose two static analyses, origin and capability analysis and error
    source analysis, to resolve the discrepancies between the I/O APIs of libc
    and Rust std (\cref{sec:analysis}).
  \item
    We propose program transformation techniques that replace the I/O API by
    assigning correct types to streams and propagating errors using the analysis
    results (\cref{sec:transformation}).
  \item
    We realize the proposed approach as a tool, Forcrat ({\bf F}\C{ILE*} {\bf
    O}peration {\bf R}eplacing {\bf C}-to-{\bf R}ust {\bf A}utomatic {\bf
    T}ranslator),
    and evaluate it on 62 real-world C programs, showing that the approach is
    (1) correct, with all 32 programs that have test suites passing the tests
    after transformation,
    (2) efficient, analyzing and transforming 422k LOC in 14 seconds, and
    (3) widely applicable, replacing 82\% of the API function calls
    (\cref{sec:evaluation}).
\end{itemize}
We also discuss related work (\cref{sec:related}) and conclude
(\cref{sec:conclusion}).

%% file: background.tex
\section{Background}
\label{sec:background}

\subsection{I/O API of libc}
\label{sec:background:c}

In libc, two different levels of I/O APIs exist:
a high-level one using streams of the \C{FILE*} type, and a low-level one using
file descriptors, expressed as integers.
This work focuses on the former because it is more widely used due to its
flexibility and convenience, while the latter is recommended for use only when
specifically necessary~\cite{glibc}.

The functions performing stream-based I/O operations are defined in the
\C{stdio.h} header file.
The POSIX 2024 standard~\cite{posix2024} specifies 66 API functions to be
declared in this file.
However, among them, 6 functions are pure string operations and unrelated to I/O
(e.g., \C{sprintf}),
and 3 functions operate on file descriptors rather than streams (e.g.,
\C{dprintf}).
For this reason, this work focuses on replacing the remaining 57 functions, such
as \C{fopen}, \C{fread}, \C{fwrite}, and \C{fseek}.
The full list is provided in the supplementary material~\cite{supp}.
We now describe the types of streams and operations on them.

\paragraph{Origins}

While streams can have various origins, they all have the same type \C{FILE*}.
Possible origins include \emph{standard input/output/error}, \emph{files}
(opened by \C{fopen}), and \emph{pipes} (created by \C{popen}).
A pipe represents standard input or output of a subprocess, where output is
selected when the second argument to \C{popen} is \CT{"}\C{r}\CT{"}, and input
when it is \CT{"}\C{w}\CT{"}.
The code below is well-typed, as streams have the same type:

\begin{lstlisting}[language=C]
FILE *f; f = stdin; f = stdout; f = stderr;
f = fopen(..); f = popen(..);
\end{lstlisting}

\paragraph{Buffering}

Each stream can be either buffered or unbuffered.
By default, standard input/output, files, and pipes are buffered, while standard
error is unbuffered.

\paragraph{Capabilities}

Streams can also have various capabilities, but the type does not distinguish
them either.
Capabilities include \emph{reading} data from the stream, \emph{writing} data to
the stream, and \emph{seeking} positions in the stream.
The following C code is well-typed, although some operations may fail at run
time because the stream does not have the capability:

\begin{lstlisting}[language=C]
FILE *f = ..;
fread(.., f); fwrite(.., f); fseek(f, ..);
\end{lstlisting}

\paragraph{Closing Streams}

Each stream can be closed by passing it to \C{fclose} (for non-pipes) or
\C{pclose} (for pipes).
Using a closed stream causes the operation to fail at run time.

\paragraph{Error Checking}

For error checking, each stream internally maintains two indicators:
an \emph{end-of-file (EOF) indicator} and an \emph{error indicator}.
When an operation on a stream fails, it sets the EOF indicator if the failure
was due to reaching the end of the stream; otherwise, it sets the error
indicator.
Calling \C{feof} on a stream checks whether the EOF indicator is set, and
\C{ferror} checks the error indicator.

\subsection{I/O API of Rust std}
\label{sec:background:rust}

\paragraph{Origins}

Rust std provides different types for streams of different origins:
\C{Stdin}, \C{Stdout}, and \C{Stderr} for standard input/output/error;
\C{File} for files;
and \C{ChildStdin} and \C{ChildStdout} for pipes.
Assigning a stream to a variable of a different origin results in a type error:

\begin{lstlisting}[language=Rust]
let f: Stdin = stdin(); // ok
f = stdout(); // type error
\end{lstlisting}

\paragraph{Buffering}

Rust std's streams can also be buffered or unbuffered.
As in libc, standard input and output are buffered, and standard error is
unbuffered.
However, files and pipes are unbuffered.
Instead, Rust std provides the \C{BufReader} and \C{BufWriter} types, which wrap
streams with buffers.
For example, \C{BufReader}\CT{<}\C{File}\CT{>} is a readable, buffered file.

\paragraph{Capabilities}

Different capabilities are represented with different \emph{traits}, each of
which represents a set of types that provide certain functionalities.
Traits resemble type classes in Haskell and serve a similar role to abstract
classes or interfaces in object-oriented languages.
In Rust terminology, a type \C{A} is said to \emph{implement} a trait \C{T} if
\C{A} belongs to \C{T}.

To represent capabilities, Rust std provides \C{Read} for reading, \C{Write} for
writing, and \C{Seek} for seeking.
\C{Read} is implemented by \C{Stdin}, \C{File}, \C{ChildStdout}, and
\C{BufReader};
\C{Write} is implemented by \C{Stdout}, \C{Stderr}, \C{File}, \C{ChildStdin},
and \C{BufWriter}.
\C{Seek} is implemented by \C{File}, and also by \C{BufReader} and \C{BufWriter}
if the underlying stream implements \C{Seek}.

In addition to the aforementioned traits, Rust std also provides the \C{BufRead}
trait for buffered reading.
It represents the capability of reading the contents of a buffer, which is
useful for implementing operations where the number of bytes to read is
determined by the data itself, e.g., when translating \C{fscanf}.
\C{BufRead} is implemented by \C{Stdin} and \C{BufReader}.
%

Since a trait is not a type per se, Rust provides two keywords to use a trait as
a type: \C{impl} and \C{dyn}.
For a trait $T$, \C{impl} $T$ and \C{dyn} $T$ are types, and $T$ is called
the \emph{bound} of these types.
While similar, \C{impl} and \C{dyn} are used for different purposes.

First, \C{impl} can be used only as the type of a function parameter to accept a
value of any type that satisfies the bound.
Passing an argument of an incorrect type results in a type error:

\begin{lstlisting}[language=Rust]
fn foo(f: impl Read) { f.read(..); }
foo(stdin()); // ok
foo(stdout()); // type error
\end{lstlisting}

\noindent
The use of \C{impl} does not incur any run-time overhead because functions with
\C{impl} are monomorphized at compile time, emitting separate versions for each
type into the binary.

When using \C{impl}, it is also possible to specify multiple bounds for a single
type using the \CT{+} symbol:

\begin{lstlisting}[language=Rust]
fn foo(f: impl Read+Seek) { f.read(..);f.seek(..); }
\end{lstlisting}

\noindent
This requires the argument to implement both traits.

On the other hand, \C{dyn} can be used anywhere to define a variable that can
store a value of any type that satisfies the bound.
It incurs run-time overhead because the value is associated with a virtual
method table, and utilizing its capabilities (i.e., calling methods) requires
virtual dispatch.
Furthermore, since values of different types may have different sizes, a \C{dyn}
type must be wrapped in a pointer type, such as \C{Box}, which represents a
pointer to a heap-allocated object:

\begin{lstlisting}[language=Rust]
let f: Box<dyn Read> = Box::new(stdin()); // ok
f = Box::new(stdout()); // type error
\end{lstlisting}

While \C{dyn} does not allow specifying multiple bounds, we can define a
\emph{subtrait} that has multiple \emph{supertraits} and use this new trait as a
bound.
Only types that implement all the supertraits can implement the subtrait, and
the subtrait inherits all the functionalities of the supertraits:

\begin{lstlisting}[language=Rust]
trait ReadSeek : Read + Seek {}
let f:Box<dyn ReadSeek> = ..; f.read(..);f.seek(..);
\end{lstlisting}

\paragraph{Closing Streams}

Rust std does not provide functions to close streams, and each stream is closed
automatically by its destructor when it is \emph{dropped}, i.e., when it is
passed to the \C{drop} function or goes out of scope.
Rust's ownership type system prohibits any use of dropped values, preventing the
use of a closed stream at compile time:

\begin{lstlisting}[language=Rust]
let f: File = File::open(..); drop(f);
f.read(..); // type error
\end{lstlisting}

Note that dropping a pointer to a stream, rather than the stream itself, does
not necessarily close it.
Raw pointers (unsafe pointers) and references (safe pointers) merely
\emph{borrow} the pointee, and dropping them does not close the stream.
However, a \C{Box} \emph{owns} the pointee, and dropping it closes the stream.
For this reason, if a variable is only intended to use a stream without closing
it, it should be defined as a borrowing pointer to prevent unintended closure.
For example, \C{f1} is a reference in the following code:

\begin{lstlisting}[language=Rust]
let f: File = File::open(..);
let f1: &mut File = &mut f; f1.read(..); drop(f1);
f.read(..); // ok
\end{lstlisting}

\paragraph{Error Checking}

Each operation on a stream returns a \C{Result} value, which is either \C{Ok}
containing the operation's result or \C{Err} containing the reason for the
failure.
The \C{Ok} and \C{Err} cases can be distinguished using \emph{pattern matching},
and in the \C{Err} case, inspecting the reason reveals whether the failure was
due to reaching the end of the stream or another issue.

%% file: analysis.tex
\section{Static Analysis}
\label{sec:analysis}

In this section, we present two static analysis techniques:
origin and capability analysis (\cref{sec:analysis:oricap}) and error source
analysis (\cref{sec:analysis:error}).
We also discuss the reasons for classifying certain streams as unsupported,
which prevents them from being replaced with Rust std streams in the subsequent
transformation phase (\cref{sec:analysis:unsupported}).
%

\subsection{Origin and Capability Analysis}
\label{sec:analysis:oricap}

This analysis aims to determine the origins and capabilities of each
\emph{location} storing a stream, where a location is a variable or a struct
field of type \C{FILE*}.
The analysis results allow the transformation to assign a single Rust type to
each location.
Since each location's type is fixed throughout the entire program, the analysis
is flow-insensitive;
it computes the set of all origins possibly stored in each location and the set
of all capabilities required by the operations performed on each location.
We formally define a location $l$ as follows:
\[
  \small
  x\in\textrm{Variables}\qquad
  t\in\textrm{Structs}\qquad
  f\in\textrm{Fields}\qquad
  l\ ::=\ x\ |\ t.f
\]

In the analysis, an origin is either $\ostdin$, $\ostdout$, $\ostderr$,
$\ofile$, or $\opipe$, and a capability is either $\cread$, $\cbufread$,
$\cwrite$, $\cseek$, or $\cclose$.
Although closing a stream is not represented by a trait in Rust std, we treat it
as a capability because the transformation must decide whether to use an owning
or borrowing type based on whether the location closes the stream.

We now define the syntax of a program being analyzed, focusing only on the key
features relevant to the analysis:
\[
  \small
  \begin{array}{@{}r@{~}r@{~}l@{}}
    p &::=& x\ |\ e.f \\
    e &::=& p\ |\ {\tt stdin}\ |\ {\tt stdout}\ |\ {\tt stderr}\ |\ {\tt fopen}()\ |\
    {\tt popen}() \\
    s &::=& p=e\ |\ {\tt fread}(p)\ |\ {\tt fscanf}(p)\ |\ {\tt fwrite}(p) \\
    &|& {\tt fseek}(p)\ |\ {\tt fclose}(p)\ |\ {\tt pclose}(p) \\
  \end{array}
\]
A \emph{path} $p$ is either a variable or a field projection, representing a
left value of an assignment or an argument to a function call.
An \emph{expression} $e$ is either a path, standard input/output/error, a file
open, or a pipe construction, representing a right value of an assignment.
A \emph{statement} $s$ is either an assignment or a call to an API function that
takes a stream.
A program is a sequence of statements;
we ignore control structures as the analysis is flow-insensitive.

We finally describe the \emph{constraints} generated by each statement.
In constraints, we use $\oriset{l}$ to denote the set of origins of a location
$l$ and $\capset{l}$ for the set of capabilities.
In addition, $\loc{p}$ denotes the location referred to by a path $p$, defined
as follows, where $\struct{e}$ is the struct type of $e$:
\[ \small \loc{x} = x \qquad \loc{e.f} = \struct{e}.f \]
Assigning a stream-creating expression to a path generates constraints regarding
the origins based on the assigned value:
\[
  \small
  \begin{array}{@{}rcl@{}}
    p={\tt stdin}  &\rightsquigarrow&\ostdin \in\oriset{\loc{p}} \\
    p={\tt stdout} &\rightsquigarrow&\ostdout\in\oriset{\loc{p}} \\
    p={\tt stderr} &\rightsquigarrow&\ostderr\in\oriset{\loc{p}} \\
    p={\tt fopen}()&\rightsquigarrow&\ofile  \in\oriset{\loc{p}} \\
    p={\tt popen}()&\rightsquigarrow&\opipe  \in\oriset{\loc{p}} \\
  \end{array}
\]
Passing a path to an API function generates constraints regarding the
capabilities based on the called function:
\[
  \small
  \begin{array}{@{}rcl@{}}
    {\tt fread}(p) &\rightsquigarrow&\cread   \in\capset{\loc{p}} \\
    {\tt fscanf}(p)&\rightsquigarrow&\cbufread\in\capset{\loc{p}} \\
    {\tt fwrite}(p)&\rightsquigarrow&\cwrite  \in\capset{\loc{p}} \\
    {\tt fseek}(p) &\rightsquigarrow&\cseek   \in\capset{\loc{p}} \\
    {\tt fclose}(p)&\rightsquigarrow&\cclose  \in\capset{\loc{p}} \\
    {\tt pclose}(p)&\rightsquigarrow&\cclose  \in\capset{\loc{p}} \\
  \end{array}
\]
The full list of capabilities required by each API function is provided in the
supplementary material~\cite{supp}.
The most interesting case is assignment from one path to another:
\[
  \small
  p_1\!=\!p_2\rightsquigarrow
  \oriset{\loc{p_2}}\!\subseteq\!\oriset{\loc{p_1}},
  \capset{\loc{p_1}}\!\subseteq\!\capset{\loc{p_2}}
\]
Since all streams stored at $p_2$ are also stored at $p_1$, each member of
$\oriset{\loc{p_2}}$ must belong to $\oriset{\loc{p_1}}$.
On the other hand, streams stored at $p_2$ must provide all the capabilities
required by those stored at $p_1$, so $\capset{\loc{p_2}}$ must contain each
member of $\capset{\loc{p_1}}$.
This is analogous to subtyping in object-oriented languages:
a subtype has more capabilities (methods) than a supertype, and a subtype can be
assigned to a supertype, but not vice versa.

Consider the following example with locations \C{x}, \C{y}, and \C{z}:

\begin{lstlisting}[language=C]
x = fopen(); fseek(x); y = stdin;
if (b) { z = x; } else { z = y; } fread(z);
\end{lstlisting}

\noindent
The analysis collects the following constraints for this program:
$\ofile\in\oriset{\C{x}}$,
$\cseek\in\capset{\C{x}}$,
$\ostdin\in\oriset{\C{y}}$,
$\oriset{\C{x}}\subseteq\oriset{\C{z}}$,
$\capset{\C{z}}\subseteq\capset{\C{x}}$,
$\oriset{\C{y}}\subseteq\oriset{\C{z}}$,
$\capset{\C{z}}\subseteq\capset{\C{y}}$,
and $\cread\in\capset{\C{z}}$.
Solving these constraints results in the following sets:
$\oriset{\C{x}}=\{\ofile\}$, $\capset{\C{x}}=\{\cseek,\cread\}$,
$\oriset{\C{y}}=\{\ostdin\}$, $\capset{\C{y}}=\{\cread\}$,
$\oriset{\C{z}}=\{\ofile,\ostdin\}$, and $\capset{\C{z}}=\{\cread\}$.

Since all constraints have the form of set membership or a subset relation, we
can solve them using the well-known cubic algorithm~\cite{spa}.
It constructs a graph representing the constraints and then updates it to
propagate the constraints across nodes by following the edges.
The algorithm has a time complexity of $O(n^3)$, where $n$ is the number of
locations.
In theory, if a target program contains a large number of locations that store
streams, the analysis may not scale well.
However, in practice, even large programs use streams in only a portion of the
codebase, and the number of analyzed locations remains relatively small, making
this analysis practically efficient.

\subsection{Error Source Analysis}
\label{sec:analysis:error}

This analysis aims to find all I/O API function calls that serve as the sources
of errors checked by each call to \C{feof} or \C{ferror}.
For each error-checking call, we independently perform a dataflow analysis to
identify the sources by starting from the call statement and traversing the
program backward.
Before describing the analysis algorithm in detail, we first present examples to
illustrate the intuition behind the analysis.

First, the analysis must handle interprocedural dataflow because error checking
and its sources may not reside in the same function.
For example, sources can appear in callees:

\begin{lstlisting}[language=C]
void foo() { bar(f); if (ferror(f)) { .. } }
void bar(FILE *g) { fread(g); }
\end{lstlisting}

\noindent
In this example, \C{foo} calls \C{bar}, which performs a read operation on the
stream, and then checks for an error.
Thus, the checking occurs in \C{foo}, while the source is the \C{fread} call in
\C{bar}.

It is also possible for sources to be in callers:

\begin{lstlisting}[language=C]
void foo(FILE *f) { if (ferror(f)) { .. } }
void bar() { FILE *g = ..; fread(g); foo(g); }
\end{lstlisting}

\noindent
Here, \C{foo} can be viewed as a custom error-checking function, and \C{bar}
calls \C{foo} after reading from the stream.
Therefore, the \C{fread} call in \C{bar} is a source for the \C{ferror} call in
\C{foo}.

Second, the analysis can stop moving backward from callees to callers if at
least one source is found.
Consider below:

\begin{lstlisting}[language=C]
void foo(FILE *f) { if (cond) { fread(f); }
                    if (ferror(f)) { .. } }
\end{lstlisting}

\noindent
This function reaches the \C{ferror} call without performing any I/O operations
if \C{cond} is false.
However, the intent of \C{ferror} is presumably to check whether \C{fread}
succeeded, not to detect errors that occurred before entering \C{foo}.
When an operation on a stream fails, the next operation on the same stream is
likely to fail again for the same reason.
For this reason, C programmers rarely perform operations on a stream whose
indicator may be already set.
Therefore, if at least one source is found before reaching the function entry,
our analysis does not continue into the callers.

We now describe the analysis algorithm, shown in \cref{alg:error}.
The algorithm \textit{findSources} identifies the sources of an error-checking
call at a label $L_0$.
We assume that each statement has a distinct label $L$.
For simplicity, we consider only variables and single-parameter functions;
the logic can be easily extended to handle struct fields and multiple
parameters.
The argument \textit{visitedCallSites} is initially an empty set and is updated
when the algorithm is called recursively.
The algorithm consists of two phases:
(1) finding sources in the current function and its callees
(lines~\ref{line:error1}--\ref{line:error20}), and
(2) finding sources in the callers
(lines~\ref{line:error21}--\ref{line:error31}).

\SetStartEndCondition{ }{}{}
\SetKwProg{Fn}{def}{\string:}{}
\SetKwFor{While}{while}{:}{fintq}
\SetKwFor{For}{for}{\string:}{}
\SetKwIF{If}{ElseIf}{Else}{if}{:}{else if}{else:}{}
\AlgoDontDisplayBlockMarkers\SetAlgoNoEnd\SetAlgoNoLine

\makeatletter
\patchcmd{\@algocf@finish}
  {\@algoskip} 
  {\@algoskip \vspace{-2em}}
  {}{}
\makeatother

\begin{algorithm}[t]
  \caption{Error source analysis}
  \label{alg:error}
  \setstretch{0.8}
  \footnotesize
  \Fn{findSources($L_0$, visitedCallSites)}{
    $x_0:=$ the argument of the function call at $L_0$\label{line:error1}\;
    \embox{worklist} $=\{(x_0,L_0)\}$; \embox{visited} $=\emptyset$; \embox{sources} $=\emptyset$\label{line:error2}\;
    \While{worklist $\not=\emptyset$\label{line:error3}}{
      $(x,L):=$ \embox{worklist}.\embox{pop}()\label{line:error4}\;
      \lIf{$(x,L)\in$ visited}{
        {\bf continue}\label{line:error6}
      }
      \embox{visited} $:=$ \embox{visited} $\cup\ \{(x,L)\}$\label{line:error7}\;
      $s:=$ the statement at $L$\label{line:error8}\;
      \If{$s=$ \embox{func}$(x)$ for some \embox{func}\label{line:error9}}{
        \If{\embox{func} is a source\label{line:error10}}{
          \embox{sources} $:=$ \embox{sources} $\cup\ \{L\}$\label{line:error11}\;
          {\bf continue}\label{line:error12}\;
        }
        \ElseIf{\embox{func} is a user-defined function\label{line:error13}}{
          $x':=$ the parameter of \embox{func}\label{line:error14}\;
          $L':=$ the label of the return statement of \embox{func}\label{line:error15}\;
          \embox{worklist} $:=$ \embox{worklist} $\cup\ \{(x',L')\}$\label{line:error16}\;
        }
        \lElseIf{\embox{func} is a function ptr}{
          \Return{$\emptyset$}\label{line:error18}
        }
      }
      \For{$L'\leftarrow$ the predecessors of $L$\label{line:error19}}{
        \embox{worklist} $:=$ \embox{worklist} $\cup\ \{(x,L')\}$\label{line:error20}\;
      }
    }
    \embox{func} $:=$ the function containing $L_0$\label{line:error21}\;
    \If{sources $=\emptyset\land x_0$ {\scriptsize is a param} $\land\ $\embox{func}
    {\scriptsize is not used as a function ptr}\label{line:error22}}{
      \embox{visitedCallSites} $:=$ \embox{visitedCallSites} $\cup\ \{L_0\}$\label{line:error23}\;
      \For{$L'\leftarrow$ the callsites of \embox{func}\label{line:error24}}{
        \If{$L'\not\in$ visitedCallSites\label{line:error26}}{
          \embox{sources}$'$ $:=$ \embox{findSources}($L'$, \embox{visitedCallSites})\label{line:error27}\;
          \lIf{sources$'=\emptyset$}{
            \Return{$\emptyset$}\label{line:error29}
          }
          \embox{sources} $:=$ \embox{sources} $\cup$ \embox{sources}$'$\label{line:error30};
        }
      }
    }
    \Return{sources}\label{line:error31}\;
  }
\end{algorithm}

In the first phase, we start by initializing \textit{worklist}, which contains
pairs of a variable and a label to be visited, \textit{visited}, which stores
already-visited pairs, and \textit{sources}, which collects the source labels
(lines~\ref{line:error1}--\ref{line:error2}).
Then, we iteratively visit each label until \textit{worklist} becomes empty
(line~\ref{line:error3}).
At each iteration, we pop a pair from \textit{worklist}, check whether it has
already been visited, and if not, add it to \textit{visited}
(lines~\ref{line:error4}--\ref{line:error7}).

If the statement at the current label is a function call on the variable
(lines~\ref{line:error8}--\ref{line:error9}), we handle three cases:
\begin{itemize}
  \item
    If the callee is an API function that performs a failable operation, we add
    the label to \textit{sources} and do not proceed to the preceding
    statements, as the source along this execution path has now been found
    (lines~\ref{line:error10}--\ref{line:error12}).
  \item
    If the callee is a user-defined function, we add the parameter name and the
    callee's return label to \textit{worklist} to search for sources inside the
    callee (lines~\ref{line:error13}--\ref{line:error16}).
  \item
    If the callee is a function pointer, the algorithm returns an empty set,
    signifying failure to find sources (line~\ref{line:error18}).
    We do not propagate errors across function pointer calls because this
    requires changing the parameter/return types of all functions that can be
    assigned to this function pointer variable, even if they are unrelated to
    I/O.
    Failing to find sources makes the stream unsupported (see
    \cref{sec:analysis:unsupported}).
    %
    %
    %
\end{itemize}
Finally, we add all the predecessors to \textit{worklist} to continue moving
backward, determining the predecessors of $L$ from the function's control flow
graph (lines~\ref{line:error19}--\ref{line:error20}).
We store $x$ in \textit{worklist} because we track both the visited label and
the variable under consideration when visiting a label.
This is necessary since multiple variables may store a stream.

The second phase begins by deciding whether we need to proceed to the callers
(lines~\ref{line:error21}--\ref{line:error22}).
If no sources have been found yet and $x_0$ is a parameter, we proceed to
analyze the callers.
However, if the current function is used as a function pointer, we return the
empty set due to the aforementioned design decision regarding function pointers.

For each call site of the function, we recursively call \textit{findSources}
(lines~\ref{line:error23}--\ref{line:error27}).
During this process, we track the set of visited call sites to ensure
termination, even when analyzing recursive functions.
If source finding fails at some call site, the overall process is considered to
have failed, and the algorithm returns an empty set
(line~\ref{line:error29}).
Otherwise, we accumulate all discovered sources and return them
(lines~\ref{line:error30}--\ref{line:error31}).

We apply \textit{findSources} to each error-checking call to identify its
sources.
With memoization of the results from each run of the algorithm, the overall
analysis has a time complexity of $O(n^2)$, where $n$ is the size of the
program, as \textit{findSources} is invoked at $O(n)$ call sites, and each
visits up to $O(n)$ labels.
In practice, the analysis often terminates in linear time because sources are
typically located close to the error-checking calls.

\subsection{Unsupported Streams}
\label{sec:analysis:unsupported}

We now discuss the reasons for classifying certain locations as unsupported.
Such locations retain the \C{FILE*} type even after the transformation.
The majority of the reasons for being unsupported stems from the limited
functionality provided by Rust std compared to libc.
We summarize 11 reasons below.

\paragraph{Setbuf}

A stream is passed to the \C{setbuf} or \C{setvbuf} function, which changes the
buffering mode of the stream.
However, Rust std does not support changing the buffering mode of standard input
and output to unbuffered.

\paragraph{Ungetc}

A stream is passed to the \C{ungetc} function, which pushes a byte into a
readable stream.
However, \C{Read} and \C{BufRead} in Rust std do not provide such functionality.

\paragraph{Freopen}

A stream is passed to the \C{freopen} function, which changes the underlying
object to a specified file.
Rust std does not support this behavior.

\paragraph{Improper Capabilities}

The capabilities required of a stream do not match its possible origins.
For example, in libc, seeking on standard input may succeed at run time if the
input has been redirected to a file;
in Rust std, however, \C{Stdin} never implements the \C{Seek} trait, regardless
of redirection.

\paragraph{Cyclic Close Capability}

The capabilities of multiple streams form a \emph{cyclic} dependency while
including $\cclose$. The following code is an example:

\vspace{-0.25em}
\begin{lstlisting}[language=C]
x = y; .. y = x; .. fclose(x);
\end{lstlisting}
\vspace{-0.25em}

\noindent
Here, the capabilities of \C{x} and \C{y} form a cycle, as they are assigned to
each other, and $\cclose$ is required due to the \C{fclose} call.
While this works in C, Rust's ownership type system allows each stream to be
owned by a single variable, preventing this form of cyclic ownership.

\paragraph{Comparison}

Two \C{FILE*} pointers are compared using \CT{==} or \CT{!=}.
Rust std does not support comparing streams.

\paragraph{Casts}

A stream is cast to or from a \C{void} pointer or an integer.
Since Rust std's streams are not pointers, they cannot be cast to or from
\C{void} pointers or integers.

\paragraph{Variadic}

A stream is passed as a variadic argument.
Variadic arguments are not type-checked, so it is meaningless to assign Rust
std's types to them.

\paragraph{API Functions as Pointers}

A stream is passed to a function pointer call that may refer to an I/O API
function from libc.
API functions such as \C{feof} and \C{ferror} cannot be represented as function
pointers using Rust std.

\paragraph{Error Sources Unfound}

A stream is passed to \C{feof} or \C{ferror}, but error source analysis failed
to find the sources.

\paragraph{Non-POSIX}

The stream is passed to a non-POSIX API function, such as \C{\_\_freading} in
GNU libc.
Non-POSIX APIs are outside the scope of this work.

\vspace{0.5em}

The unsupported-ness of a location propagates through assignments,
\emph{bidirectionally}, from the right value to the left value and vice versa.
For example, in the following code, if \C{y} is unsupported, then not only \C{x}
but also \C{z} become unsupported:

\begin{lstlisting}[language=C]
if (..) { x = y; } else { x = z; }
\end{lstlisting}

\noindent
This is because we cannot convert a libc stream to or from a Rust std stream at
run time.
If \C{z} is a Rust std stream while \C{y} is a libc stream, they cannot be
assigned to the same variable.

Since the unsupported-ness of the left value and right value of an assignment is
the same, we can compute the complete set of unsupported locations using the
unification algorithm based on the union-find data
structure~\cite{galler1964improved}.
This runs in $O(n \cdot \alpha(n))$ time complexity, where $\alpha$ is the
inverse Ackermann function.

%% file: transformation.tex
\section{Program Transformation}
\label{sec:transformation}

We now present program transformation techniques that replace libc streams in
C2Rust-generated code with Rust std streams using the analysis results.
The transformation consists of four steps:
transforming types (\cref{sec:transformation:types}),
transforming stream construction expressions
(\cref{sec:transformation:streams}),
transforming assignments and function calls
(\cref{sec:transformation:assignments}),
and propagating errors (\cref{sec:transformation:errors}).
%

\subsection{Transforming Types}
\label{sec:transformation:types}

In this step, we replace each type annotation of \C{*mut\ FILE} (Rust syntax for
\C{FILE*}) with an appropriate type from Rust std.
We define the stream types provided by Rust std as follows:
\[
  \small
  \begin{array}{@{}r@{~}r@{~}l@{}}
    B &::=& \tstdin\ |\ \tstdout\ |\ \tstderr\ |\ \tfile \\
    &|& \tbufreader\ |\ \tbufwriter\ |\ \tchild \\
    T &::=& \tread\ |\ \tbufread\ |\ \twrite\ |\ \tseek \\
    \beta &::=& \overline{T} \\
    \tau &::=& B\ |\ \tptr{B}\ |\ \tbox{\beta}\ |\ \tdynptr{\beta}\ |\
    \timpl{\beta} \\
  \end{array}
\]

$B$ is a \emph{base type}, where {\small$\tstdin$}, {\small$\tstdout$},
{\small$\tstderr$}, {\small$\tfile$}, and {\small$\tchild$} correspond to each
origin, and {\small$\tbufreader$} and {\small$\tbufwriter$} wrap files with
buffers.
We use \C{Child} for pipes instead of \C{ChildStdin} or \C{ChildStdout}, due to
the behavior of \C{pclose} in libc, which waits for the subprocess to terminate.
In Rust std, a subprocess is represented by \C{Child}, which can be waited on
and contains \C{ChildStdin} and \C{ChildStdout} as fields for I/O.
To ensure correct translation of \C{pclose}, we use \C{Child} as a stream
originating from a pipe.

$T$ is a trait corresponding to each capability except $\cclose$.
$\beta$ is a list of traits, representing bounds.

A type $\tau$ is either a base type, a raw pointer to a base type, \C{dyn}
wrapped by \C{Box}, \C{dyn} wrapped by a raw pointer, or \C{impl}.
%
%
%
We use raw pointers instead of references because references require their
lifetimes to be decided, which is unrelated to I/O and outside the scope of this
work.
Existing techniques~\cite{emre2021translating, emre2023aliasing,
zhang2023ownership} can be applied to replace raw pointers with references.

\begin{algorithm}[t]
  \caption{Type decision}
  \label{alg:type}
  \setstretch{0.8}
  \footnotesize
  \Fn{decideType($l$)}{
    \lIf{$l$ is a parameter}{
      \Return{$\timpl{\boundof{\capset{l}}}$}\label{line:type2}
    }\ElseIf{$\#(\oriset{l})=1$}{\label{line:type3}
      \lIf{$\oriset{l}=\{\ostdin\}$}{
        $B:=\tstdin$\label{line:type4}
      }\lElseIf{$\oriset{l}=\{\ostdout\}$}{
        $B:=\tstdout$\label{line:type5}
      }\lElseIf{$\oriset{l}=\{\ostderr\}$}{
        $B:=\tstderr$\label{line:type6}
      }\ElseIf{$\oriset{l}=\{\ofile\}$}{\label{line:type7}
        \lIf{$\cwrite\not\in\capset{l}$}{
          $B:=\tbufreader$\label{line:type8}
        }\ElseIf{$\{\cread,\cbufread\}\cap\capset{l}=\emptyset$}{\label{line:type9}
          $B:=\tbufwriter$\;\label{line:type10}
        }\lElse{
          $B:=\tfile$\label{line:type11}
        }
      }\lElse{
        $B:=\tchild$\label{line:type12}
      }
      \lIf{$\cclose\in\capset{l}$}{
        \Return{$B$}\label{line:type13}
      }\lElse{
        \Return{$\tptr{B}$}\label{line:type14}
      }
    }\Else{\label{line:type15}
      \lIf{$\cclose\in\capset{l}$}{
        \Return{$\tbox{\boundof{\capset{l}}}$}\label{line:type16}
      }\lElse{
        \Return{$\tdynptr{\boundof{\capset{l}}}$}\label{line:type17}
      }
    }
  }
\end{algorithm}

\cref{alg:type} shows how we decide the type of each location $l$ based on its
origins and capabilities.
The type determined by this algorithm replaces the type annotation of the
location.
We treat the cases where $l$ is a parameter and where it is not differently.
We first discuss the non-parameter cases, which are further divided into
single-origin and multi-origin cases.

When only one origin is possible, we determine the base type according to the
origin (lines~\ref{line:type4}--\ref{line:type12}).
A notable case is when the origin is $\ofile$, where we also examine the
capabilities.
If the location is not used for writing, we use {\small$\tbufreader$}, and if it
is not used for reading, we use {\small$\tbufwriter$}, making the stream
buffered.
However, if the location is used for both reading and writing, we use
{\small$\tfile$} to support both kinds of operations.
Finally, if the location is used for closing, we assign the base type as the
type of the location (line~\ref{line:type13});
otherwise, we use a raw pointer (line~\ref{line:type14}).

When multiple origins are possible, we use \C{dyn} to allow storing different
base types in the location.
If the location is used for closing, we wrap the \C{dyn} type in a \C{Box}
(line~\ref{line:type16});
otherwise, we use a raw pointer (line~\ref{line:type17}).
Here, ${\it bound}$ converts the given set of capabilities into bounds by
mapping each capability to its corresponding trait, while ignoring $\cclose$.

Finally, when $l$ is a parameter, we use \C{impl} regardless of the number of
origins to give the most general type to the parameter (line~\ref{line:type2}).
We also do not check whether the stream is used for closing.
This is because we can handle this at the call site instead---passing an owning
value to close the stream, or a borrowing value to leave it open.
%


In fact, since libc streams can be null pointers while Rust std streams cannot,
we need to use \C{Option}\CT{<}$\tau$\CT{>} instead of $\tau$ as the type of
each location.
An \C{Option} value is either \C{Some(}$v$\C{)}, representing a non-null stream,
or \C{None}, representing a null stream.
However, this issue of nullability arises not only in the context of I/O, but
also when handling any raw pointer;
previous studies that replace raw pointers with references also wrap references
in \C{Option}~\cite{emre2021translating, emre2023aliasing, zhang2023ownership}.
Thus, the remaining discussion omits \C{Option} to focus on I/O-specific
aspects.

\subsection{Transforming Stream Construction Expressions}
\label{sec:transformation:streams}

In this step, we transform each expression that constructs a libc stream into an
expression that constructs a Rust std stream.
We replace \C{stdin}, \C{stdout}, and \C{stderr} with \C{stdin()}, \C{stdout()},
and \C{stderr()} because standard streams are global variables in libc but are
returned from functions in Rust std.
We also replace \C{fopen} calls with calls to functions that return \C{File},
such as \C{File{::}open}, and replace \C{popen} calls with
\C{Command{::}new(..).spawn()}, which returns \C{Child}.

\subsection{Transforming Assignments and Function Calls}
\label{sec:transformation:assignments}

In this step, we transform assignments and function calls involving streams.
This is necessary because the types of the left value (or parameter) and the
right value (or argument) may differ.
For example, the left value can have more origins than the right value according
to the analysis, resulting in the left value having \C{Box} while the right
value has a base type.
Therefore, we must transform each right value to make it conform to the type of
the left value.

\begin{table}[t]
  \caption{Transformation of right values according to types}
  \label{tab:rvalue}
  \vspace{-1em}
  \centering
  $
  \footnotesize
  \begin{array}{@{~}c@{~}||@{~}c@{~}|@{~}c@{~}|@{~}c@{~}|@{~}c@{~}}
    & B_r & \tptr{B_r} & \tbox{\beta_r} & \tdynptr{\beta_r} \\ \hline\hline
    B_l & e\ ^\dagger & & & \\ \hline
    \tptr{B_l} & \eptr{e} & e & & \\ \hline
    \tbox{\beta_l} & {\tt Box{::}new}(e) & & e & \\ \hline
    \tdynptr{\beta_l} & \eptr{e} & e & e.{\tt as\_mut\_ptr}() & e \\
  \end{array}
  $
  \vspace{0.25em}

  {\scriptsize ${}^\dagger$
  ${\tt BufReader{::}new}(e)$ when $B_l=\tbufreader$ and $B_r=\tfile$;\\[-0.5em]
  ${\tt BufWriter{::}new}(e)$ when $B_l=\tbufwriter$ and $B_r=\tfile$}
  \vspace{-1em}
\end{table}

We first discuss the transformation of the right values of assignments, as
summarized in \cref{tab:rvalue}.
Rows represent the types of left values, and columns represent the types of
right values.
Each cell shows how the right value is transformed, where $e$ denotes the
original expression.
Since \C{impl} types only appear as parameter types when transforming function
calls, they are not considered in assignment transformations.

An empty cell in the table indicates that such a case never occurs.
Since the right value has more capabilities than the left, we never convert a
borrowing value to an owning value.
Likewise, as the right value has fewer origins, we never convert a \C{dyn} type
to a base type.

During the transformation, we only compare the categories of the types (i.e.,
base type, raw pointer, \C{Box\ dyn}, and \C{dyn} pointer) and do not check
whether the inner base types ($B_l$ and $B_r$) or bounds ($\beta_l$ and
$\beta_r$) are compatible.
This is because types that make conversion infeasible never occur, as types are
determined from the origins and capabilities computed by the analysis.
For example, we never transform {\small$\tstdin$} to {\small$\tstdout$}.
The only exception is when converting {\small$\tfile$} to {\small$\tbufreader$}
or {\small$\tbufwriter$}, where we use \C{BufReader{::}new(}$e$\C{)} or
\C{BufWriter{::}new(}$e$\C{)}, instead of $e$.

\begin{table}[t]
  \caption{Transformation of arguments according to types}
  \label{tab:arg}
  \vspace{-1em}
  \centering
  $\footnotesize
  \begin{array}{@{~}c@{~}|@{~}c@{~}|@{~}c@{~}|@{~}c@{~}}
    B & \tptr{B} & \tbox{\beta} & \tdynptr{\beta} \\ \hline\hline
    {\tt\&mut\ }e & e & e.{\tt as\_mut}() & e \\
  \end{array}$
  \vspace{-1.5em}
\end{table}

We now describe the transformation of arguments for user-defined function calls.
If the parameter of the function requires $\cclose$, then the argument remains
unchanged, as its type is guaranteed to be a base type or \C{Box}, and passing
the owning value allows the function to close the stream.
On the other hand, if the parameter does not require $\cclose$, we transform the
argument based on its type, as shown in \cref{tab:arg}.
For owning values, we take their addresses to prevent the stream from being
closed;
for borrowing values, we pass them as-is.

Finally, we discuss the transformation of libc API function calls.
The only two functions that close a stream are \C{fclose} and \C{pclose}.
We transform these into calls to \C{drop}, keeping the arguments unchanged.
Other function calls are translated into their Rust std equivalents.
For example, \C{fputc(}$v, e$\C{)} becomes $e'$\C{.write\_all(\&[}$v$\C{])}, where
$e'$ is the transformed form of $e$ according to \cref{tab:arg}.
The supplementary material~\cite{supp} lists all libc I/O functions and their
translations in Rust std.

Below is an example showing the result of the transformation up to this step:

\begin{lstlisting}[language=Rust]
// Before transformation
let x: *mut FILE = fopen(..);
let y: *mut FILE = popen(..);
let z: *mut FILE;
if cond { z = x; } else { z = y; }
fputc('a', z); fclose(x); pclose(y);
// After transformation
let x: BufWriter<File> = BufWriter::new(File(..));
let y: Child = Command::new(..).spawn();
let z: *mut dyn Write;
if cond { z = &raw mut x; } else { z = &raw mut y; }
z.write_all(&['a']); drop(x); drop(y);
\end{lstlisting}

\noindent
Here, \C{x} and \C{y} have base types because each has a single origin and is
closed, while \C{z} is a \C{dyn} pointer because it has multiple origins and is
not closed, necessitating proper conversions in the assignments from \C{x} and
\C{y} to \C{z}.

\subsection{Propagating Errors}
\label{sec:transformation:errors}

In this step, we transform the program to propagate errors.
We first define a local variable to store a stream indicator in each function
that either checks the indicator or contains an error source.
Each variable is initialized to \C{0}, indicating no error.
We also replace each \C{feof}/\C{ferror} call with an expression that reads this
error variable.

We then transform each Rust std API call acting as a source to check whether the
returned \C{Result} is \C{Err}.
If so, we update the error variable to a non-zero value, indicating an error.

Finally, we transform user-defined functions to propagate errors.
If a source is in a callee, we modify the callee to return a tuple consisting of
the original return value and the error value;
at the call site, we use the returned error value to update the caller's error
variable.
Conversely, if a source is in a caller, we transform the callee to take an
additional parameter, and modify the caller to pass the error variable as an
argument.

Below is an example showing the result of the transformation, where \C{bar}
contains a source:

\begin{lstlisting}[language=Rust]
// Before transformation
fn bar(x:*mut FILE)->i32 {fputc('a', x); return ..;}
bar(y); if ferror(y) != 0 { .. }
// After transformation
fn bar(x: impl Write) -> (i32, i32) { let e = 0;
    if x.write_all(&['a']).is_err() { e = 1; }
    return (.., e); }
let e=0; let (_, _e)=bar(&mut y); e=_e; if e!=0 {..}
\end{lstlisting}

Unfortunately, the transformed code does not follow Rust's idioms for error
handling.
%
The idiomatic approach would be to transform functions to take or return
\C{Option}/\C{Result} values instead of integers.
Since this work focuses on preserving semantics during I/O translation, we leave
improving idiomaticity to future work.

%% file: evaluation.tex
\section{Evaluation}
\label{sec:evaluation}

In this section, we evaluate our approach using 62 real-world C programs.
We first describe our implementation of Forcrat, which realizes the proposed
approach (\cref{sec:evaluation:implementation}), the benchmark programs used
for evaluation (\cref{sec:evaluation:benchmark}), and the overall experimental
process (\cref{sec:evaluation:process}).
The implementation and benchmarks are publicly available~\cite{artifact}.
We then address the following research questions:
\begin{itemize}
  \item \textbf{RQ1. Correctness}: Does it transform programs while preserving their
    semantics? (\cref{sec:evaluation:correctness})
  \item \textbf{RQ2. Efficiency}: Does it analyze and transform programs efficiently?
    (\cref{sec:evaluation:efficiency})
  \item \textbf{RQ3. Applicability}: Is it applicable to replacing a wide range of libc
    I/O API calls? (\cref{sec:evaluation:applicability})
  \item \textbf{RQ4. Impact on performance}: What is the effect of the transformation on
    program performance? (\cref{sec:evaluation:performance})
  \item \textbf{RQ5. Security}: How much does the transformation improve the
    security of programs? (\cref{sec:evaluation:security})
  \item \textbf{RQ6. Code changes}: How much does the code change due to the
    transformation? (\cref{sec:evaluation:code})
\end{itemize}
Our experiments were conducted on an Ubuntu machine with an Intel Core i7-6700K
(4 cores, 8 threads, 4GHz) and 32GB of DRAM.
Finally, we discuss threats to validity (\cref{sec:evaluation:threats}).

\subsection{Implementation}
\label{sec:evaluation:implementation}

We built Forcrat on top of the Rust compiler~\cite{rustc}.
For static analysis, we utilize Rust's mid-level intermediate representation
(MIR)~\cite{mir}, which expresses functions as control flow graphs composed of
basic blocks.
For program transformation, we modify the abstract syntax tree and then
pretty-print it back to code.
We used C2Rust v0.18.0 with minor modifications.

\subsection{Benchmark Programs}
\label{sec:evaluation:benchmark}

\renewcommand{\arraystretch}{0.45}
\begin{table}[t]
  \caption{Benchmark programs}
  \label{tab:benchmark}
  \centering
\scriptsize
  \vspace{-0.5em}
  \begin{tabular}{@{~}l@{~}||@{~}r@{~}|@{~}r@{~}|@{~}r@{~}|@{~}r@{~}}
    \textbf{Name} & \textbf{C LOC} & \textbf{Rust LOC} & \textbf{API Calls} & \textbf{Remaining Calls} \\
    \hline
avl&101&114&1&0\\
bc-1.07.1&10,810&16,982&158&36\\
brotli-1.0.9&13,173&127,691&99&0\\
bst&65&89&1&0\\
buffer-0.4.0&395&1,137&9&0\\
bzip2&5,316&13,731&226&87\\
cflow-1.7&20,601&26,375&150&11\\
compton&8,748&14,084&130&3\\
cpio-2.14&35,934&80,929&163&2\\
dap-3.10&22,420&43,549&1,391&27\\
diffutils-3.10&59,377&95,835&236&99\\
ed-1.19&2,439&5,636&66&6\\
enscript-1.6.6&34,868&78,749&987&234\\
findutils-4.9.0&80,015&139,859&311&38\\
gawk-5.2.2&58,111&140,566&643&517\\
genann-1.0.0&608&1,818&69&0\\
glpk-5.0&71,805&145,738&82&28\\
gprolog-1.5.0&52,193&74,381&583&80\\
grep-3.11&64,084&84,902&70&11\\
gsl-2.7.1&227,199&422,756&880&0\\
gzip-1.12&20,875&21,605&97&0\\
hello-2.12.1&8,340&10,688&33&11\\
heman&7,048&14,690&12&0\\
hiredis&7,305&14,042&481&0\\
indent-2.2.13&19,255&15,581&125&29\\
less-633&20,063&45,685&64&18\\
libcsv&965&3,010&9&0\\
libosip2-5.3.1&15,772&36,286&134&0\\
libtool-2.4.7&3,769&5,701&8&7\\
libtree-3.1.1&1,412&2,632&149&0\\
libzahl-1.0&2,438&4,096&1&0\\
lil&2,934&5,558&28&0\\
lodepng&5,098&14,299&11&0\\
make-4.4.1&28,911&36,336&362&1\\
mcsim-6.2.0&18,527&36,454&906&78\\
minilisp&722&2,149&24&7\\
mtools-4.0.43&18,266&37,021&627&10\\
nano-7.2&42,999&74,995&102&25\\
nettle-3.9&61,835&82,742&290&0\\
patch-2.7.6&28,215&103,839&207&41\\
pexec-1.0rc8&5,357&12,301&140&0\\
pocketlang&14,267&41,439&82&0\\
pth-2.0.7&7,590&12,950&178&9\\
quadtree-0.1.0&365&1,057&14&0\\
raygui&1,588&17,218&38&0\\
rcs-5.10.1&28,286&36,267&106&99\\
rgba&396&2,128&2&0\\
robotfindskitten&398&1,508&3&0\\
screen-4.9.0&39,335&72,201&113&3\\
sed-4.9&48,190&68,465&426&266\\
shairport&4,995&10,118&66&3\\
tar-1.34&66,172&134,970&305&80\\
time-1.9&2,828&1,830&102&0\\
tinyproxy&5,667&12,825&56&0\\
tulipindicators&12,371&22,864&45&0\\
twemproxy&22,738&74,593&199&0\\
units-2.22&7,240&11,521&241&44\\
urlparser&56&1,360&12&0\\
uucp-1.07&51,123&77,872&1,150&297\\
webdis&14,369&29,474&24&5\\
wget-1.21.4&81,188&192,742&325&125\\
which-2.21&2,010&2,241&42&0\\\hline
{\bf Total}&&&13,594&2,337\\
  \end{tabular}
  \vspace{-1em}
\end{table}

We used benchmark programs from previous studies on improving C2Rust-generated
code~\cite{zhang2023ownership, hong2024dont, hong2024tag}.
This resulted in 62 programs, including 36 GNU packages as well as open-source
projects from GitHub.
Columns 2--4 of \cref{tab:benchmark} present the C LOC, Rust LOC, and the number
of libc I/O API calls in each program, respectively.
The benchmarks are diverse in size, ranging up to 422,756 Rust LOC and 1,391 API
calls.
Among these, 32 programs are accompanied by test suites.

\subsection{Experimental Process}
\label{sec:evaluation:process}

We first translated all benchmark programs using C2Rust.
We checked whether the translated programs were compilable and found that 25
programs were not, due to type errors.
We manually fixed these errors, modifying an average of 11.7 lines per program.
We also ran the test suites, when available, and confirmed that all tests
passed.
Next, we applied our approach to the translated code, performing static analysis
and code transformation while measuring the times to answer RQ2.
After this, we checked whether the transformed programs were compilable and
passed their test suites to answer RQ1.
We also measured the runtime of each test suite to answer RQ4.
Finally, we inspected the transformed code to count the number of transformed
API calls and to measure the amount of unsafe code and code changes, addressing
RQ3, RQ5, and RQ6.

\subsection{RQ1: Correctness}
\label{sec:evaluation:correctness}

We evaluate the correctness of the proposed approach by checking whether the
transformed program is compilable and passes its test cases.
First, the experimental results show that our approach produces compilable code
for most programs.
Among the 62 programs, only one program, nano-7.2, fails to compile.
We manually investigated this case and confirmed that the failure is not due to
a problem in our approach.
Below is the problematic part in the original C code:

\begin{lstlisting}[language=C]
if (original != NULL) { v = copy_file(original); }
if (original == NULL || v < 0) { .. }
\end{lstlisting}

\noindent
Here, \C{original} is a stream, and \C{copy\_file} closes it.
Although the program uses \C{original} even after it is closed, it is only to
check nullity, not to perform I/O, which is not considered a bug.
However, Forcrat translates this into the following uncompilable Rust code:

\begin{lstlisting}[language=Rust]
if !original.is_none() { v = copy_file(original); }
if original.is_none() || v < 0 { .. }
\end{lstlisting}

\noindent
Since Rust's ownership type system prevents the use of a variable after it is
passed to another function, using \C{original} after calling \C{copy\_file}
results in a type error.
To obtain compilable code through translation, we can fix the original program
as shown below, which checks nullity and stores the result in a variable before
calling \C{copy\_file}:

\begin{lstlisting}[language=C]
int is_original_null = original == NULL;
if (!is_original_null) { v = copy_file(original); }
if (is_original_null || v < 0) { .. }
\end{lstlisting}

This example suggests that our approach can affect programs in ways not directly
related to I/O operations.
In C, programs can use a variable that stores a stream (e.g., for nullity
checking) even after closing it.
In contrast, Rust allows the use of only owned variables, and closed stream
variables are no longer considered owned.
As a result, any statements that access closed stream variables can be affected
by the transformation, potentially making the program uncompilable.
This implies that developers may need to slightly revise their code before
translation to satisfy Rust's strict ownership discipline.
In addition, although this specific case is not a bug, it supports the common
belief that translating to Rust can reveal previously unknown bugs, such as
using a stream for I/O operations after it has been closed.

The experimental results also show that our transformation likely preserves
program behavior.
All the 32 programs still pass the tests after the transformation.

To further investigate correctness, we conducted case studies.
To effectively evaluate correctness, we need to focus on the possible sources of
incorrect transformation that arise from the design of our approach.
An incorrect transformation can result from:
(1) incorrect origin and capability analysis or an incorrect decision on stream
types,
(2) incorrect transformation of stream construction, assignment, or function
call expressions, or
(3) incorrect error propagation.

Among these, the first two do not require manual checking.
First, an incorrect stream type results in a type error and is automatically
detected by the compiler.
Second, we transform stream construction, assignment, and user-defined function
call expressions only to match the types of the left and right values, and any
incorrect transformation also leads to a type error.
In addition, since the transformation of API function calls relies on the same
mappings across all programs, it is meaningless to perform case studies on
individual translations.

Therefore, our case studies focus on the correctness of error propagation.
Since each error-checking call is independently transformed, each case study
targets a single error-checking call.
Among the 62 programs, we found 140 transformed error-checking calls across 24
programs.
In each program, we studied all calls if it contained three or fewer, and
randomly sampled three otherwise, resulting in 59 calls studied in total.
The authors and another Rust expert independently reviewed each call, and we
confirmed that their assessments agreed.

Our studies reveal that the error propagation of 8 calls (1 from bc-1.07.1, 2
from gprolog-1.5.0, 1 from nettle-3.9, 2 from patch-2.7.6, 1 from time-1.9, and
1 from uucp-1.07) is potentially incorrect.
The reason is that these programs use error-checking calls to detect errors from
multiple API calls, rather than a single call, as illustrated below:

\begin{lstlisting}[language=C]
fputc('a', x); fputc('b', x); if (ferror(x)) { .. }
\end{lstlisting}

\noindent
Since our analysis considers only the API call found first while moving backward
as the source, earlier calls are not treated as sources, and the indicator value
is not propagated from them.
Therefore, the transformed code can behave incorrectly when an earlier call
fails but a later call succeeds.
We believe this issue is rarely problematic in practice, because if an earlier
call fails, a later call is likely to fail again.
Nevertheless, addressing this issue would be an interesting direction for future
work.

\subsection{RQ2: Efficiency}
\label{sec:evaluation:efficiency}

\begin{figure}[t]
  \centering
  \includegraphics[width=0.78\columnwidth]{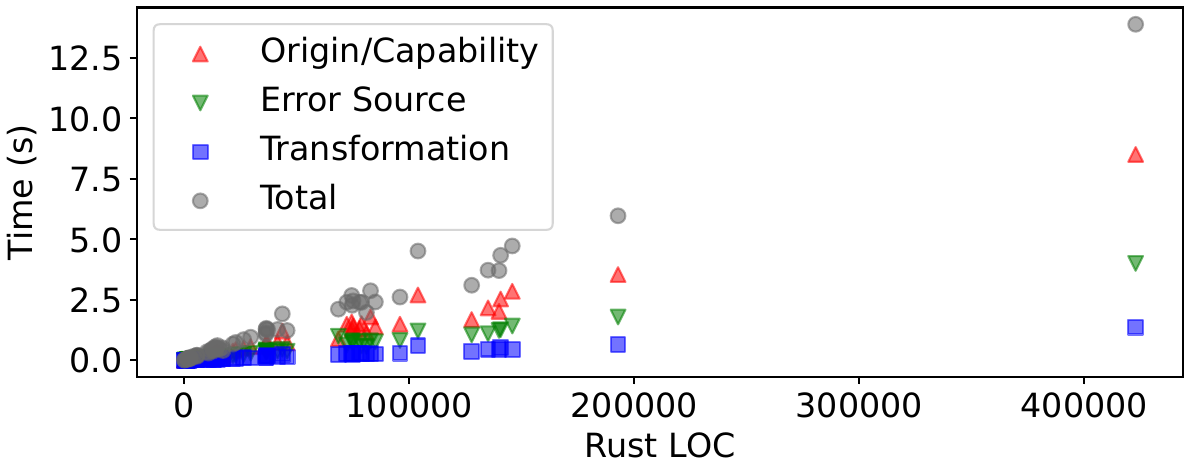}
  \vspace{-1em}
  \caption{Forcrat execution time}
  \vspace{-1.5em}
  \label{fig:time}
\end{figure}

We evaluate the efficiency of the proposed approach by measuring the time taken
to analyze and transform each program.
The experimental results show that our approach is highly efficient.
\cref{fig:time} presents the origin and capability analysis time, error source
analysis time, transformation time, and the total time, all relative to the Rust
LOC.
Even for the largest program, gsl-2.7.1 with 422k LOC, the entire process takes
less than 14 seconds.
Note that the time required to translate each benchmark program using C2Rust
ranges from 0.6 to 295 seconds.
On average, running the proposed method on C2Rust's output takes only 8.9\% of
the time spent by C2Rust, demonstrating the efficiency of the proposed approach.

The results provide empirical evidence that the proposed approach scales well to
large programs in practice.
Although the origin and capability analysis has cubic time complexity in theory,
it exhibited nearly linear performance in our experiments.
This is because real-world programs contain a relatively small number of
stream-storing locations compared to their overall size.
The majority of analysis time is spent collecting constraints by visiting each
statement, rather than solving the constraints.
To confirm this, we also measured the constraint-solving time and found that it
completed in less than 5 milliseconds for each program.

Similarly, the error source analysis is quadratic in theory, but it also showed
nearly linear performance in our experiments.
This is because error-checking calls and their sources are typically located
close to each other, allowing the source to be found after analyzing only nearby
functions.
We measured the number of functions analyzed per error-checking call and found
that only 2.15 functions were analyzed on average.

\subsection{RQ3: Applicability}
\label{sec:evaluation:applicability}

We evaluate the applicability of the proposed approach by counting the number of
libc I/O API calls replaced in each program.
If a stream is classified as unsupported by the analysis, its type remains
\C{*mut\ FILE} during transformation, and the I/O API calls on that stream are
not replaced with Rust std API calls.
The practical value of our approach is high only if it can replace the majority
of API calls.
Therefore, we assess whether our approach is applicable to a sufficiently large
number of API calls.

The experimental results show that our approach achieves high applicability by
replacing 11,257 out of 13,594 libc I/O API calls (82\%) across all benchmark
programs.
The last column of \cref{tab:benchmark} shows the number of libc I/O API calls
that remain after the transformation.
Notably, in 28 out of the 62 programs, Forcrat successfully replaces all API
calls.
We believe these results are promising, especially considering the design
differences between C and Rust.
For example, \textsc{Crown}~\cite{zhang2023ownership} replaces 30\% of raw
pointers with references in a subset of these benchmark programs, highlighting
the inherent difficulty of translating C features into Rust.

Nevertheless, future research is needed to replace more API calls.
For example, Forcrat exhibits a low conversion ratio in a few programs:
below 50\% for four programs (gawk-5.2.2, libtool-2.4.7, rcs-5.10.1, and
sed-4.9).

\begin{figure}[t]
  \centering
  \includegraphics[width=0.9\columnwidth]{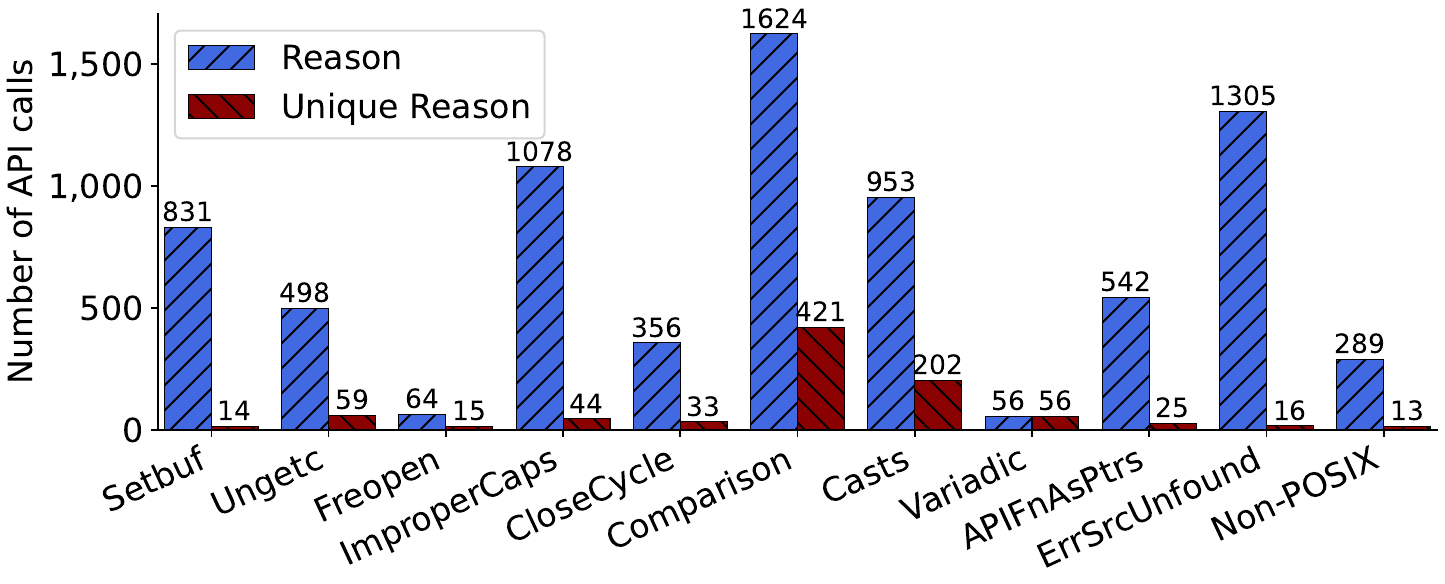}
  \vspace{-1em}
  \caption{Reasons for failing to replace API calls}
  \vspace{-1.5em}
  \label{fig:reason}
\end{figure}

To guide future research directions, we investigate the reasons for failing to
replace each API call, which correspond to the reasons for classifying the
stream used as the argument of the call as unsupported.
We follow the 11 reasons described in \cref{sec:analysis:unsupported} in
this investigation.
Note that a single call can have multiple reasons, e.g., when the argument
stream is passed to \C{setbuf} and also used in a pointer comparison.
\cref{fig:reason} presents the results:
blue bars show the number of API calls affected by each reason, and red bars
show the number of calls affected \emph{solely} by that unique reason.

The results show that many calls are affected by \emph{multiple} reasons, so
resolving a single reason alone would not lead to significant improvement.
That said, Comparison and Casts appear as the unique reason for more than a
hundred calls, making them higher-priority targets for future work.
To address Comparison, it may be feasible to compare the underlying file
descriptors of streams, as Rust std supports obtaining file descriptors from
streams.
For Casts, since Wu and Demsky's work~\cite{wu2025genc2rust} replaces \C{void}
pointers with Rust generics, adopting their approach could handle such cases.
In addition, we may reduce the number of streams that are unsupported due to
Improper Capabilities by developing a flow-sensitive version of the origin and
capability analysis.
If a stream variable has different origins and capabilities at different program
locations, a flow-sensitive analysis can capture this information, allowing us
to split the variable into multiple variables with different types.

\subsection{RQ4: Impact on Performance}
\label{sec:evaluation:performance}

We evaluate the impact of replacing libc streams with Rust std streams on
program performance.
We first investigate how many \C{dyn}-type variables are introduced by the
transformation, as \C{dyn} incurs run-time overhead due to virtual dispatch.

It turns out that \C{dyn} variables are rarely introduced---only 0.9 per program
on average.
In 48 programs, no \C{dyn} variables are introduced at all.
The maximum occurs in cflow-1.7, where 14 \C{dyn} variables are introduced.
We believe this is because C programmers typically handle streams from different
origins by passing them to the same function, but rarely by storing them in the
same variable.

In addition, we compare the performance of each Rust program before and after
the transformation by measuring the execution time of the test suite.
To ensure reliable results, we used only the 25 programs whose test suite
execution time exceeds 0.1 seconds and ran each program fifty times.

The experimental results show that the performance difference is negligible.
The transformed programs were slower by only 1.23\% on average compared to the
original ones.
In 9 programs, execution was even faster after the transformation.
We also performed a one-sided Welch's t-test on the results from each program
with the null hypothesis that the transformed program is slower than the
original by 5\%.
With a significance level of 0.05, we could reject the null hypothesis for 15
out of 25 programs, implying that the transformation is unlikely to incur
significant performance overhead.
While we could not reject the null hypothesis for the remaining 10 programs,
this does not necessarily mean that the overhead exceeds 5\%.
Increasing the number of runs may yield stronger statistical significance,
allowing us to reject the null hypothesis.

\subsection{RQ5: Security}
\label{sec:evaluation:security}

We evaluate how much the proposed approach improves the security of programs.
In Rust, code security is often approximated by measuring the amount of
\emph{unsafe} code because the compiler cannot verify its safety and leaves the
burden of validation to the developers.
Following this convention, we measured the number of characters in the unsafe
code before and after the transformation.

The results show that the transformation slightly improves security by reducing
unsafe code by 4.4\%.
This reduction comes from replacing API calls, as all libc API calls are
considered unsafe by the Rust compiler, while most Rust std API calls are
considered safe.
Unfortunately, the security improvement is small because unsafe code arises from
various sources~\cite{emre2021translating}.
Achieving a significant improvement requires addressing diverse features and
APIs, and our work represents an important step toward this long-term goal.


\subsection{RQ6: Code Changes}
\label{sec:evaluation:code}

We evaluate how much the transformation changes the code by counting the line
differences between the code before and after the transformation.
This helps estimate the effort required to manually replace the I/O API, and
also
provides insight into the verbosity of the transformed code.

The results show that the transformation relieves a significant burden but
increases code verbosity.
On average, 1,904 lines are inserted and 553 lines are deleted per program.
Manually modifying the code to this extent would require substantial effort.
At the same time, this indicates that each line is replaced with roughly four
lines.
This increase in code size is mainly due to
(1) some libc API functions being replaced with sequences of Rust std API calls,
owing to the lack of exact equivalents, and
(2) error propagation requiring the introduction and updates of error variables.
Reducing the verbosity would be an interesting direction for future research.

\subsection{Threats to Validity}
\label{sec:evaluation:threats}

Threats to external validity stem from the choice of benchmarks, which consist
of GNU packages and open-source projects on GitHub.
Since GNU packages often share common code patterns, including many GNU packages
in the benchmarks may introduce bias.
Although open-source projects enhance diversity, they still cannot fully
represent the entire C ecosystem.
Expanding the experiments to include a wider variety of C programs would
strengthen the evidence for the generalizability of our approach.

Threats to construct validity arise from using test suites for correctness.
Since tests do not cover all behaviors of the programs, passing all tests does
not guarantee semantics preservation.
However, we discovered bugs in our implementation during development, indicating
that the test suites are reasonably effective.
That said, testing the transformed programs more thoroughly using fuzzing or
automated unit test generation would provide stronger evidence of correctness.

Threats to construct validity arise also from using test suites for performance
comparison.
The test suites are not designed for performance measurement, and many exhibit
high variance in execution time.
To mitigate this, we ran each program multiple times and used statistical tests.
Furthermore, we discovered improper API mappings that led to performance
degradation during development, indicating that the test suites are suitable for
revealing performance issues.
However, developing dedicated benchmarks for performance evaluation would
facilitate a stronger analysis of performance impact.

%% file: related.tex
\section{Related Work}
\label{sec:related}

\subsection{Improving C2Rust-Generated Code with Static Analysis}

Several previous studies focus on improving C2Rust-generated code using static
analysis.
Concrat~\cite{hong2023concrat} is the most relevant to this work;
it replaces libc's lock API with Rust std's lock API.
To resolve the discrepancies between the two lock APIs, it performs static
analysis to identify the data protected by each lock and the locks held at each
program point.
Other studies target language features other than library functions.
\textsc{Laertes}~\cite{emre2021translating, emre2023aliasing} and
\textsc{Crown}~\cite{zhang2023ownership} replace raw pointers with references by
analyzing lifetimes and ownership, while GenC2Rust~\cite{wu2025genc2rust}
replaces \C{void} pointers with generics by computing typing constraints.
In addition, Urcrat~\cite{hong2024tag} replaces unions with tagged unions, and
Nopcrat~\cite{hong2024dont} replaces output parameters with tuples and
\C{Option}s.

\subsection{C-to-Rust Translation via Large Language Models}

Recent studies have explored using large language models (LLMs) to translate C
to Rust.
While LLMs are proficient at generating readable and concise code, their
translations are often incorrect, resulting in uncompilable code or altered
semantics, even for small programs such as competitive programming solutions.
Hong and Ryu~\cite{hong2024type} used GPT-4o mini and observed that 44\% of
functions became uncompilable;
%
%
Yang et al.~\cite{yang2024vert} used Claude 2 and reported that 35\% of
functions were uncompilable and 25\% failed property-based tests.
To improve correctness, Shetty et al.~\cite{shetty2024syzygy} guided LLMs using
pointer information collected via dynamic analysis, while Zhang et
al.~\cite{zhang2024scalable} provided hand-written feature mappings for
Go-to-Rust translation.
An interesting direction would be to guide LLMs using the static analysis
proposed in this work.

\subsection{Static Analysis of Code Using I/O APIs}

Previous studies propose typestate analysis that can verify the correct use of
I/O APIs.
\textsc{CQual}~\cite{foster2002type} extends C with flow-sensitive type
qualifiers;
Plaid~\cite{sunshine2011first} supports first-class typestates;
Fink et al.'s framework~\cite{fink2008effective} performs typestate verification
for Java.
Typestate analysis verifies flow-sensitive properties, e.g., a stream should not
be read after being closed.
In contrast, our origin and capability analysis focuses on flow-insensitive
properties, e.g., whether a certain location is used for closing, to determine
correct types for transformation.
After the transformation, the Rust type checker can verify some flow-sensitive
properties because the notion of ownership in Rust is flow-sensitive.
For example, reading a closed stream corresponds to using a variable without
ownership in Rust, which results in a type error.

%% file: conclusion.tex
\section{Conclusion}
\label{sec:conclusion}

In this work, we tackle the problem of replacing the I/O API in C2Rust-generated
code using static analysis.
To address the different sets of types representing origins and capabilities, we
propose origin and capability analysis;
to address the different error-checking mechanisms, we propose error source
analysis.
Our evaluation shows that the proposed approach is correct, efficient, and
widely applicable.

%% file: paper.bbl
\begin{thebibliography}{10}
\providecommand{\url}[1]{#1}
\csname url@samestyle\endcsname
\providecommand{\newblock}{\relax}
\providecommand{\bibinfo}[2]{#2}
\providecommand{\BIBentrySTDinterwordspacing}{\spaceskip=0pt\relax}
\providecommand{\BIBentryALTinterwordstretchfactor}{4}
\providecommand{\BIBentryALTinterwordspacing}{\spaceskip=\fontdimen2\font plus
\BIBentryALTinterwordstretchfactor\fontdimen3\font minus
  \fontdimen4\font\relax}
\providecommand{\BIBforeignlanguage}[2]{{%
\expandafter\ifx\csname l@#1\endcsname\relax
\typeout{** WARNING: IEEEtran.bst: No hyphenation pattern has been}%
\typeout{** loaded for the language `#1'. Using the pattern for}%
\typeout{** the default language instead.}%
\else
\language=\csname l@#1\endcsname
\fi
#2}}
\providecommand{\BIBdecl}{\relax}
\BIBdecl

\bibitem{chen2011linux}
\BIBentryALTinterwordspacing
H.~Chen, Y.~Mao, X.~Wang, D.~Zhou, N.~Zeldovich, and M.~F. Kaashoek, ``{Linux}
  kernel vulnerabilities: state-of-the-art defenses and open problems,'' in
  \emph{Proceedings of the Second Asia-Pacific Workshop on Systems}, ser. APSys
  '11.\hskip 1em plus 0.5em minus 0.4em\relax New York, NY, USA: Association
  for Computing Machinery, 2011. [Online]. Available:
  \url{https://doi.org/10.1145/2103799.2103805}
\BIBentrySTDinterwordspacing

\bibitem{msrcblog}
G.~Thomas, ``A proactive approach to more secure code,''
  \url{https://msrc-blog.microsoft.com/2019/07/16/a-proactive-approach-to-more-secure-code},
  2019.

\bibitem{matsakis2014rust}
\BIBentryALTinterwordspacing
N.~D. Matsakis and F.~S. Klock, ``The {Rust} language,'' in \emph{Proceedings
  of the 2014 ACM SIGAda Annual Conference on High Integrity Language
  Technology}, ser. HILT '14.\hskip 1em plus 0.5em minus 0.4em\relax New York,
  NY, USA: Association for Computing Machinery, 2014, p. 103–104. [Online].
  Available: \url{https://doi.org/10.1145/2663171.2663188}
\BIBentrySTDinterwordspacing

\bibitem{jung2017rustbelt}
\BIBentryALTinterwordspacing
R.~Jung, J.-H. Jourdan, R.~Krebbers, and D.~Dreyer, ``{RustBelt}: Securing the
  foundations of the {Rust} programming language,'' \emph{Proc. ACM Program.
  Lang.}, vol.~2, no. POPL, dec 2017. [Online]. Available:
  \url{https://doi.org/10.1145/3158154}
\BIBentrySTDinterwordspacing

\bibitem{rust-curl}
T.~Hutt, ``Would {Rust} secure {cURL}?''
  \url{https://blog.timhutt.co.uk/curl-vulnerabilities-rust/}, 2021.

\bibitem{li2024rust}
Z.~Li, V.~Narayanan, X.~Chen, J.~Zhang, and A.~Burtsev, ``{Rust} for {Linux}:
  Understanding the security impact of {Rust} in the {Linux} kernel,'' in
  \emph{2024 Annual Computer Security Applications Conference (ACSAC)}, 2024,
  pp. 548--562.

\bibitem{c2rust}
``{C2Rust},'' \url{https://github.com/immunant/c2rust}.

\bibitem{c2rust-huawei}
Y.~Yu, A.~d'Antras, and N.~D.~Q. Bui, ``Our {Rust} mission at {Huawei},''
  \url{https://trusted-programming.github.io/2021/02/07/our-rust-mission-at-huawei.html},
  2021.

\bibitem{zebra}
M.~Racek, ``{Zebra.rs},'' \url{https://github.com/panstromek/zebra-rs}.

\bibitem{qcms}
``{qcms},'' \url{https://github.com/FirefoxGraphics/qcms/}.

\bibitem{emre2021translating}
\BIBentryALTinterwordspacing
M.~Emre, R.~Schroeder, K.~Dewey, and B.~Hardekopf, ``Translating {C} to safer
  {Rust},'' \emph{Proc. ACM Program. Lang.}, vol.~5, no. OOPSLA, oct 2021.
  [Online]. Available: \url{https://doi.org/10.1145/3485498}
\BIBentrySTDinterwordspacing

\bibitem{emre2023aliasing}
\BIBentryALTinterwordspacing
M.~Emre, P.~Boyland, A.~Parekh, R.~Schroeder, K.~Dewey, and B.~Hardekopf,
  ``Aliasing limits on translating {C} to safe {Rust},'' \emph{Proc. ACM
  Program. Lang.}, vol.~7, no. OOPSLA1, apr 2023. [Online]. Available:
  \url{https://doi.org/10.1145/3586046}
\BIBentrySTDinterwordspacing

\bibitem{zhang2023ownership}
H.~Zhang, C.~David, Y.~Yu, and M.~Wang, ``Ownership guided {C} to {Rust}
  translation,'' in \emph{Computer Aided Verification}, C.~Enea and A.~Lal,
  Eds.\hskip 1em plus 0.5em minus 0.4em\relax Cham: Springer Nature
  Switzerland, 2023, pp. 459--482.

\bibitem{wu2025genc2rust}
\BIBentryALTinterwordspacing
X.~Wu and B.~Demsky, ``{GenC2Rust}: Towards generating generic {Rust} code from
  {C},'' in \emph{2025 IEEE/ACM 47th International Conference on Software
  Engineering (ICSE)}.\hskip 1em plus 0.5em minus 0.4em\relax Los Alamitos, CA,
  USA: IEEE Computer Society, May 2025, pp. 664--664. [Online]. Available:
  \url{https://doi.ieeecomputersociety.org/10.1109/ICSE55347.2025.00127}
\BIBentrySTDinterwordspacing

\bibitem{hong2024tag}
\BIBentryALTinterwordspacing
J.~Hong and S.~Ryu, ``To tag, or not to tag: Translating {C}'s unions to
  {Rust}'s tagged unions,'' in \emph{Proceedings of the 39th IEEE/ACM
  International Conference on Automated Software Engineering}, ser. ASE
  '24.\hskip 1em plus 0.5em minus 0.4em\relax New York, NY, USA: Association
  for Computing Machinery, 2024, p. 40–52. [Online]. Available:
  \url{https://doi.org/10.1145/3691620.3694985}
\BIBentrySTDinterwordspacing

\bibitem{hong2024dont}
\BIBentryALTinterwordspacing
------, ``Don't write, but return: Replacing output parameters with algebraic
  data types in {C}-to-{Rust} translation,'' \emph{Proc. ACM Program. Lang.},
  vol.~8, no. PLDI, jun 2024. [Online]. Available:
  \url{https://doi.org/10.1145/3656406}
\BIBentrySTDinterwordspacing

\bibitem{hong2023concrat}
\BIBentryALTinterwordspacing
------, ``{Concrat}: An automatic {C}-to-{Rust} lock {API} translator for
  concurrent programs,'' in \emph{Proceedings of the 45th International
  Conference on Software Engineering}, ser. ICSE '23.\hskip 1em plus 0.5em
  minus 0.4em\relax Melbourne, Victoria, Australia: IEEE Press, 2023, p.
  716–728. [Online]. Available:
  \url{https://doi.org/10.1109/ICSE48619.2023.00069}
\BIBentrySTDinterwordspacing

\bibitem{zhong2010mining}
\BIBentryALTinterwordspacing
H.~Zhong, S.~Thummalapenta, T.~Xie, L.~Zhang, and Q.~Wang, ``Mining {API}
  mapping for language migration,'' in \emph{Proceedings of the 32nd ACM/IEEE
  International Conference on Software Engineering - Volume 1}, ser. ICSE
  '10.\hskip 1em plus 0.5em minus 0.4em\relax New York, NY, USA: Association
  for Computing Machinery, 2010, p. 195–204. [Online]. Available:
  \url{https://doi.org/10.1145/1806799.1806831}
\BIBentrySTDinterwordspacing

\bibitem{nguyen2014statistical}
\BIBentryALTinterwordspacing
A.~T. Nguyen, H.~A. Nguyen, T.~T. Nguyen, and T.~N. Nguyen, ``Statistical
  learning approach for mining {API} usage mappings for code migration,'' in
  \emph{Proceedings of the 29th ACM/IEEE International Conference on Automated
  Software Engineering}, ser. ASE '14.\hskip 1em plus 0.5em minus 0.4em\relax
  New York, NY, USA: Association for Computing Machinery, 2014, p. 457–468.
  [Online]. Available: \url{https://doi.org/10.1145/2642937.2643010}
\BIBentrySTDinterwordspacing

\bibitem{meng2012history}
S.~Meng, X.~Wang, L.~Zhang, and H.~Mei, ``A history-based matching approach to
  identification of framework evolution,'' in \emph{Proceedings of the 34th
  International Conference on Software Engineering}, ser. ICSE '12.\hskip 1em
  plus 0.5em minus 0.4em\relax Zurich, Switzerland: IEEE Press, 2012, p.
  353–363.

\bibitem{glibc}
``The {GNU} {C} library reference manual,''
  \url{https://www.gnu.org/software/libc/manual/pdf/libc.pdf}.

\bibitem{posix2024}
``{IEEE/Open Group Standard for Information Technology--Portable Operating
  System Interface (POSIX™) Base Specifications, Issue 8},'' \emph{IEEE/Open
  Group Std 1003.1-2024 (Revision of IEEE Std 1003.1-2017)}, pp. 1--4107, 2024.

\bibitem{supp}
\BIBentryALTinterwordspacing
J.~Hong and S.~Ryu, ``{Forcrat}: Automatic {I/O} {API} translation from {C} to
  {Rust} via origin and capability analysis (supplementary material),'' Jun.
  2025. [Online]. Available: \url{https://doi.org/10.5281/zenodo.15574125}
\BIBentrySTDinterwordspacing

\bibitem{spa}
A.~M\o{}ller and M.~I. Schwartzbach, \emph{Static Program Analysis}, jun 2024,
  {Department} of Computer Science, Aarhus University,
  \url{http://cs.au.dk/\~amoeller/spa/}.

\bibitem{galler1964improved}
\BIBentryALTinterwordspacing
B.~A. Galler and M.~J. Fisher, ``An improved equivalence algorithm,''
  \emph{Commun. ACM}, vol.~7, no.~5, p. 301–303, May 1964. [Online].
  Available: \url{https://doi.org/10.1145/364099.364331}
\BIBentrySTDinterwordspacing

\bibitem{artifact}
\BIBentryALTinterwordspacing
J.~Hong and S.~Ryu, ``{Forcrat}: Automatic {I/O} {API} translation from {C} to
  {Rust} via origin and capability analysis (artifact),'' May 2025. [Online].
  Available: \url{https://doi.org/10.5281/zenodo.15559525}
\BIBentrySTDinterwordspacing

\bibitem{rustc}
Rust, ``Guide to {Rustc} development,''
  \url{https://rustc-dev-guide.rust-lang.org/}, 2024.

\bibitem{mir}
------, ``Guide to {Rustc} development: The {MIR},''
  \url{https://rustc-dev-guide.rust-lang.org/mir/index.html}, 2024.

\bibitem{hong2024type}
J.~Hong and S.~Ryu, ``Type-migrating {C}-to-{Rust} translation using a large
  language model,'' \emph{Empirical Software Engineering}, vol.~30, no.~1, Oct.
  2024.

\bibitem{yang2024vert}
\BIBentryALTinterwordspacing
A.~Z.~H. Yang, Y.~Takashima, B.~Paulsen, J.~Dodds, and D.~Kroening, ``{VERT}:
  Verified equivalent {Rust} transpilation with large language models as
  few-shot learners,'' 2024. [Online]. Available:
  \url{https://arxiv.org/abs/2404.18852}
\BIBentrySTDinterwordspacing

\bibitem{shetty2024syzygy}
\BIBentryALTinterwordspacing
M.~Shetty, N.~Jain, A.~Godbole, S.~A. Seshia, and K.~Sen, ``{Syzygy}: Dual
  code-test {C} to (safe) {Rust} translation using {LLMs} and dynamic
  analysis,'' 2024. [Online]. Available: \url{https://arxiv.org/abs/2412.14234}
\BIBentrySTDinterwordspacing

\bibitem{zhang2024scalable}
\BIBentryALTinterwordspacing
H.~Zhang, C.~David, M.~Wang, B.~Paulsen, and D.~Kroening, ``Scalable, validated
  code translation of entire projects using large language models,'' 2024.
  [Online]. Available: \url{https://arxiv.org/abs/2412.08035}
\BIBentrySTDinterwordspacing

\bibitem{foster2002type}
J.~S. Foster, \emph{Type qualifiers: lightweight specifications to improve
  software quality}.\hskip 1em plus 0.5em minus 0.4em\relax University of
  California, Berkeley, 2002.

\bibitem{sunshine2011first}
\BIBentryALTinterwordspacing
J.~Sunshine, K.~Naden, S.~Stork, J.~Aldrich, and E.~Tanter, ``First-class state
  change in {Plaid},'' in \emph{Proceedings of the 2011 ACM International
  Conference on Object Oriented Programming Systems Languages and
  Applications}, ser. OOPSLA '11.\hskip 1em plus 0.5em minus 0.4em\relax New
  York, NY, USA: Association for Computing Machinery, 2011, p. 713–732.
  [Online]. Available: \url{https://doi.org/10.1145/2048066.2048122}
\BIBentrySTDinterwordspacing

\bibitem{fink2008effective}
\BIBentryALTinterwordspacing
S.~J. Fink, E.~Yahav, N.~Dor, G.~Ramalingam, and E.~Geay, ``Effective typestate
  verification in the presence of aliasing,'' \emph{ACM Trans. Softw. Eng.
  Methodol.}, vol.~17, no.~2, may 2008. [Online]. Available:
  \url{https://doi.org/10.1145/1348250.1348255}
\BIBentrySTDinterwordspacing

\end{thebibliography}
